# STUDY OF NONLINEARITIES AND SMALL PARTICLE LOSSES IN HIGH POWER LINAC

A. Kolomiets, S. Yaramishev, ITEP, Moscow, Russia

*Abstract*

The conception of High Power Linac developed in Russian accelerator centres is based on the use of independently phased SC resonators with quadrupole lenses between them. The type and parameters of the resonators as well as focusing structure are varied along the linac to optimise beam dynamics and the characteristics of the linac. The beam evolution in the linac was studied by simulation in 3D accelerating and focusing fields by computer code DYNAMION. The simulation includes all nonlinearities of external fields and space charge forces. Estimations of particle losses in the beam based on analysis of the spectral properties of particle trajectories were carried out.

## 1 INTRODUCTION

The accelerator driven electronuclear installation for numerous purposes requires the proton beam with energy about 1 GeV and current between one and several tens milliamp [1]. The only linear resonant accelerator is considered as a choice if beam current above 10 mA is required.

In accordance with the conception proton linac will be built using one channel scheme. It consists of 0.1 MeV DC injector, room temperature 300 MHz RFQ, intermediate part with low beta 300 MHz SC independently phased cavities and main part with multigap 600 MHz SC cavities.

The paper is devoted to study of beam dynamics in the intermediate part of the linac. It is clear that perturbation of the beam in this part of the linac is the most strongly marked due to low particle velocity and consequently high influence of space charge as well as relatively high defocusing in the cavities

The intermediate part of linac has the energy range from $\beta \approx 0.15$ to $\beta \approx 0.5$. In this part the short (1 - 2 gaps) SC cavities with independent RF excitation of each resonator are the best choice as accelerating structure.

The most powerful method of the beam dynamics study is computer simulation. Many codes are used for this purpose. The output of the codes is usually set of particles coordinates and velocities stored at some structure positions. Evolution of the beam parameters is estimated by calculation of rms or total beam emittances. However the condition corresponding to the harsh emittance growth are wittingly out of acceptable range for high power linacs where the particle losses are the most critical problem. It means that the development of more sensitive methods suitable for analysis of computer simulation results is actual task.

Some new approaches to beam analysis have been proposed and studied in ITEP. In the paper the methods are described. The results of the dynamics study in low energy section of intermediate part of HPL are presented.

## 2 TRANSVERSE DYNAMICS OF THE PERTURBED BEAM

The studies initiated by interest to high power linac development and widely carried out showed that the halo formation is connected with appearance of stochastic elements in the dynamics under the influence of wide range of factors causing the perturbation of linear motion. These factors are space charge forces, influence of longitudinal motion, mismatching, etc. The appearance of stochastic elements means that certain number of the particle trajectories became similar some random function. It occurs even in the system where no random forces influence the particle motion.

It follows from general theory of non-linear dynamics [2] that the appearance of such trajectories is the result of local instabilities in the system. This process leads to the mixing of the trajectories in phase space and to the emittance increasing. If the local instability is the main reason of increasing of particle amplitudes it is sufficient to determine the conditions when it appears and find appropriate quantitative characteristic.

It is known [3] that the charged particle motion in periodic focusing and accelerating channel is described by Matieu-Hill equation. The fundamental solutions of the equation are Floquet function. The characteristic parameter of the equation determines stability or instability of the solutions.

The transformation of particle coordinate through focusing period of the channel is described be matrix $T$ with elements built from Floquet functions,

$$z(\tau+1) = Tz(\tau)$$

where z – vector of particle coordinates in phase space, $\tau$ - dimensionless time ($\tau = \frac{v}{S}t$, $v$ – particle velocity and $S$ – length of focusing period). Characteristic parameter $l$ can be found from expression:

$$\cosh l = \frac{1}{2} Sp(T) \qquad (1)$$

Real or complex values of $l = k+i\mu$, correspond to regions of instability of particle motion, imagine one $l=i\mu$ - stable regions. The particle trajectory in the stable

region is:
$$x(\tau) = A_0 \rho(\tau)\cos(\mu(\tau)\tau + \vartheta_0) \quad (2)$$

where $\rho(\tau)$ and $\mu(\tau)$ are module of Floquet function and phase advance, $A_0$ and $\theta_0$ – initial conditions. In linear theory all particle trajectories determine by the same Floquet functions and depend only of initial amplitude and phase. It can be assumed that for small perturbation of linear motion above mentioned expressions are valid, with $\rho(\tau)$ and $\mu(\tau)$ are the functions of position of the particle in phase space.

The transformation matrix and therefore Floquet functions in non-linear case can be found using code for simulation of particle motion. It calculates particle coordinates and velocities by linear transformation along integration step under influence of external and space charge forces $F_n$. The elementary transformation matrix, for example in X plane is:

$$\begin{pmatrix} x \\ \frac{dx}{d\tau} \end{pmatrix}_{n+1} = \begin{pmatrix} 1 & \Delta\tau_n \\ \frac{F_n \Delta\tau_n}{x_n} & 1 \end{pmatrix} \begin{pmatrix} x \\ \frac{dx}{d\tau} \end{pmatrix}_n,$$

where $\Delta\tau$ is step of integration. Matrix of full focusing period $T$ can be obtained by multiplication of the elementary matrixes. The characteristic parameter, module and phase of Floquet function can be easily calculated from the elements of the matrix. The distribution of these parameters can characterize the degree of perturbation of the motion. The total perturbation of the system can be expressed by averaging all instability increments $k_j>1$ over phase space:

$$h_0 = <\sum \ln(k_j)>.$$

It is shown in [2] that in assumption that average value of increment does not change in time, the increasing of the volume occupied by particles in phase space, appeared due to trajectory mixing under influence of local instability, can be determined as

$$\varepsilon(\tau) = \varepsilon_0 e^{2\langle h_0 \rangle \tau}. \quad (3)$$

The second proposed method consists in that stored with certain step particle coordinates obtained as a result of computer simulation are considered as some random set of points. The corresponding trajectory can be reconstructed using well-known correlation function method [4]. The particle trajectory is represented in this case by series

$$x_j = a_0 + \sum_m a_m \cos(\omega_m \tau), \quad \omega_m = 2\pi \frac{m}{N} \quad (4)$$

where $N$ - number of points. It allows study spectral properties of the initial random process generating the points. The application of this method to the study of spectral properties of particle motion in periodic focusing structures is described in [5]. It is shown there that in spectra of particle transverse frequencies in the presence of space charge forces always presents zero frequency peak due to appearance in certain number of particle trajectories expressed by (2) the term $a_0 \neq 0$.

This coefficient is the constant term of the trajectory. It is clear, that it can appear as the result of increasing of amplitude of particle oscillation due to local instabilities experienced by the given particle at some part of the structure. It is shown in [6] that the probability of increasing amplitude on the value $\Delta x$ can be described by Maxwell distribution

$$f(\Delta x) = 4 \cdot \frac{(\Delta x)^2}{\sqrt{\pi}\xi^3} e^{-(\frac{\Delta x}{\xi})^2} \quad (5)$$

Taking into account that the particle distributed on transverse coordinate as

$$g(x) = \frac{2}{\sigma\sqrt{2\pi}} e^{-\frac{1}{2}(\frac{x}{\sigma})^2}, \quad (6)$$

the probability for particle to increase its amplitude up to aperture value i.e. $x + \Delta x > a$ and, therefore to be lost, can be expressed as

$$p = \int_0^a g(x) \int_{a-x}^a f(z)dz\,dx \quad (7)$$

## 3. SIMULATION RESULTS

The described methods have been applied to analysis of simulation results carried out for intermediate part of high power linac. The code DYNAMION [7] with routine added for matrix coefficients calculation for each particle has been used for simulations. The parameters of the studied structure are given in Table 1.

**Table 1. Parameters of the focusing period of studied structure**

| Focusing lattice | FODO |
|---|---|
| Length of focusing period (cm) | 64.0 |
| Length of gap (cm) | 5.0 |
| Number of gap in cavity | 2 |
| Voltage in gap (kV) | 500 |
| Length of quadrupole (cm) | 12.0 |
| Aperture (cm) | 1.0 |
| Gradient (T/m) | 19.0 |

To increase statistic, the beam passed through studied period 100 times. The initial particle distribution was Gaussian. Matched Twiss parameters were calculated in smooth approximation for the envelopes at $2\sigma$ level. To avoid the influence of the possible mismatching of the initial distribution, beam preliminary passed through period 200 times. To keep the average particle velocity constant the reference particle passed each cavity at phase $-90^0$. The accelerating field was chosen that to have the design value of longitudinal oscillation frequency. The 3D distributions of the accelerating and focusing fields were used for simulations. Particle – particle interaction algorithm was used for space charge forces calculation. Table 2 shows some beam parameters calculated with

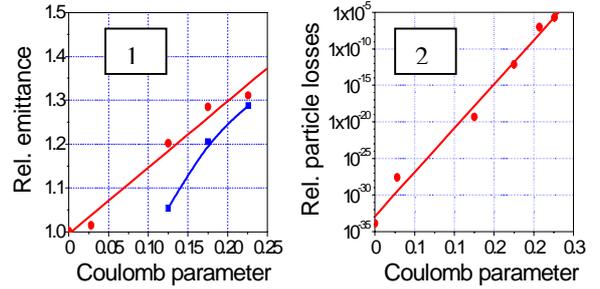

**Figure 1.** Emittance growth calculated for 100 periods (plot 1) and specific relative particle losses.

figures that there is no threshold of local instability and, therefore, of emittance growth and particle losses. The last ones are linear function of Coulomb parameter.

## 5. CONCLUSION

The proposed methods of analysis allow obtaining from results of computer simulations of beam dynamics the quantitative estimations of beam parameters, which can characterise non-linear effects. They can be useful for fast estimations with visualisation of the processes in the beam caused by non-linear motion.

The results of the work confirm that emittance growth is connected with local instability and can be estimated by its value averaged over phase space.

The study of intermediate part of conceptual proton high power linac using the methods showed that design parameters are feasible and relative particle losses at design beam current can be estimated at level $2 \cdot 10^{-6}$ per meter what is acceptable level for this part of the linac.

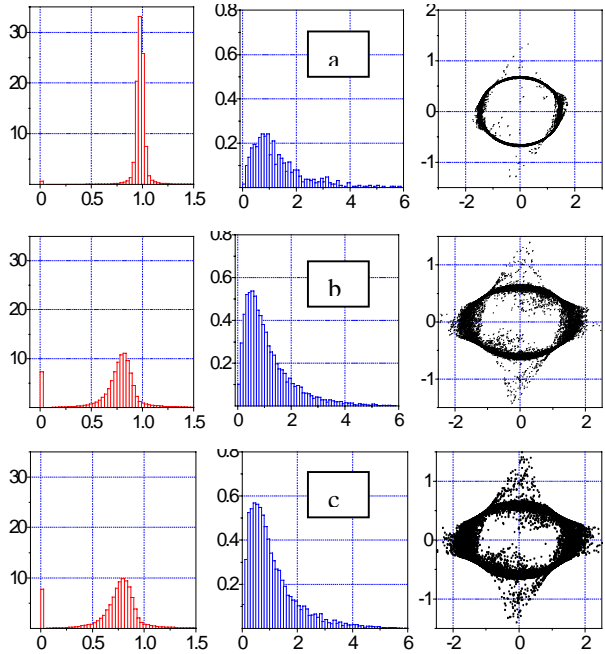

**Figure 2.** Phase advances (left column), increment of instability (centre column), normalised particle coordinates (right column) for beam currents 0 mA (a), 10 mA (b), 30 mA (c).

elements of matrices and correlation function for several values of Coulomb parameter $h \sim I/Vp$ ($I$ is beam current and $Vp$ is input normalized emittance [3]). $<\mu>$ is average phase advance, $h_0$ – average instability increment, $\sigma$ and $\xi$ are parameters of the distributions (5) and (6) correspondingly. Fig.1 shows phase advance histograms (left column of plots) and instability increment histograms (central column of plots). Right column shows phase space plots $x/A_0(\tau)$, $(dx/d\tau)/x/A_0(\tau)$ represented by module and phases of Floquet functions from (2). The rows of plots correspond to the values of Coulomb parameters 0.028, 0.175 and 0.226. It is clear seen how the increasing the number of unstable particles leads to redistribution of phase space.

The Fig.2 shows emittance growth for 100 periods (upper curve of plot 1) and probability of particle losses per meter (plot 2) in studied structure calculated from expressions (3) and (7). The lower line in Fig.2 (1) represents rms emittance of simulated beam after passing the studied period 100 times. It can be seen from given

**Table 2. Some calculated beam parameters**

| $h$ | $<\mu>$ | $h_0 \, 10^{-5}$ | $\sigma$ | $\xi \, 10^{-3}$ |
|---|---|---|---|---|
| 0.0 | 0.97 | 1.3 | 0.084 | 0.86 |
| 0.03 | 0.94 | 6.8 | 0.092 | 2.54 |
| 0.12 | 0.80 | 83.5 | 0.110 | 3.78 |
| 0.17 | 0.75 | 114.0 | 0.143 | 5.29 |
| 0.21 | 0.71 | 149.0 | 0.198 | 7.16 |
| 0.23 | 0.73 | 213.0 | 0.229 | 8.09 |